\documentclass[conference]{IEEEtran}
\IEEEoverridecommandlockouts
\usepackage{cite}
\usepackage{amsmath,amssymb,amsfonts}
\usepackage{algorithmic}
\usepackage{graphicx}
\usepackage{float}
\usepackage{textcomp}

\usepackage{lscape, tabularx}

\usepackage[hidelinks=true]{hyperref}
\usepackage{xcolor}

\def\BibTeX{{\rm B\kern-.05em{\sc i\kern-.025em b}\kern-.08em
    T\kern-.1667em\lower.7ex\hbox{E}\kern-.125emX}}
\begin{document}

\title{Computational pathology in renal disease: a comprehensive perspective\\}


\author{\IEEEauthorblockN{ Manuel Cossio}
\IEEEauthorblockA{\textit{Dept. of Mathematics and Computer Science} \\
\textit{Universitat de Barcelona}\\
Barcelona, Spain \\
Zürich, Switzerland \\
manuel.cossio@ub.edu}

}

\maketitle

\begin{abstract}
Computational pathology is a field that has complemented various subspecialties of diagnostic pathology over the last few years. In this article we briefly analyze the different applications in nephrology. We begin by providing an overview of the different forms of image production. We continue by describing the most frequent applications of computer vision models, the salient features of the different clinical applications, and the data protection considerations encountered. We finished the development, delving into the interpretability of these applications, expanding in depth the three dimensions of this area. 
\end{abstract}


\section{Introduction}
 Thanks to the expanding research in artificial intelligence, especially computer vision, computational pathology applications have increased considerably in recent years. 

 Digital pathology can be defined as the process or set of processes by which slides with already processed samples are digitized. Digitizing implies transferring the image that is observed in a light microscope to an image file\cite{cui2021artificial}. These files are called WSI (Whole Slide Image) and have a characteristic that makes them essential for use in pathology: they store different levels of magnification in the same file. This type of configuration is called a pyramid and it allows to optimize the visualization and exploration of the image, by allowing macroscopic exploration at very low levels of magnification and then going to the detail of interest by increasing the magnification\cite{litjens20181399, zarella2019practical, cossio_2022}.
The WSI can be stored on physical servers located in the same hospital or can be stored in cloud storage systems. Its storage has allowed the interconsultation of the same case between several pathologists from different parts of the world, considerably expanding the accuracy of the diagnosis in complex cases\cite{hanna2022integrating, kim2021literature, cossio_2022}.

 It can be said that digital pathology ends with the digitization of the slide and its storage in the cloud or the physical server\cite{cui2021artificial}. Part of the extension of the definition of digital pathology, recently includes the remote consultation of cases between several institutions from different physical locations in real time\cite{kim2021literature}.
On the other hand, computational pathology includes all those projects and/or prototypes where WSI images are used as a dataset and machine learning or artificial vision algorithms are employed\cite{abels2019computational, farris2021artificial, cossio_2022}. Computational pathology does not include the use of scanners, or microscopes, or the implementation of molecular biology laboratory procedures. In this discipline, we already have the images stored, which have the authorization of the ethics committee of the donor institutions and, if the case warrants, the informed consent of the patients to whom they belong\cite{sorell2022ethical}. Computational pathology also includes digital modification of WSIs (cropping, alignment, contrast, brightness, color channels, color normalization, etc.\cite{tellez2019quantifying}) and clinical data capture by aligning WSIs with electronic health records\cite{cossio_2022}.

\section{Data workflow}
The data workflow in digital pathology can take many forms. However, the main skeleton is more or less homogeneous in different countries, health networks, hospitals and laboratories. We can begin to describe it by naming its first part: data acquisition. As we have already seen in other publications, data acquisition begins with the capture of the image by the scanner, once the sample is treated and prepared following the determined protocol (for example, H\&E). That captured image must be temporarily stored and then reviewed by a qualified professional to determine if the quality is correct for diagnosis (focus and tissue area). This verification is usually done with the help of the barcode (linear or 2D) that identifies the sample and attaches it to the patient's file. Once the image is verified as correct (without interpreting the diagnostic), the storage of the image is moved to a permanent physical or virtual disk. If the disk is physical, the storage server may be located in the same hospital or in the same laboratory. If the server is virtual, the host system can be in the same country or in a different country. Now, at the moment of establishing the diagnosis (interpreting the image), using it for a virtual consultation (let it be interpreted by another professional pathologist in another place) or when an algorithm analyzes it autonomously, the image must be made available\cite{cossio_2022}. 

For this, a transmission of the file must be generated and a deposit of the same (copy) in another physical place or a read-only link can be generated, without the need for file exchange.

\section{Article structure}

In this article, a general review will be made of the current state of computational pathology in the field of nephrology. For this, the techniques used to produce images in renal pathology will be described first with their specific examples. Then, the most common tasks in computational pathology focused on renal pathology will be delved into. Some important details for training automation and data protection will be also analyzed. Finally, the interpretability in the field of images, computational pathology and computational pathology applied to renal pathology will be shown and discussed in depth. Therefore, the article will have the following structure:

\begin{enumerate}
    \item  \nameref{1}
    \item  \nameref{2}
    \item  \nameref{3}
    \item  \nameref{4}
    \item  \nameref{5}
    \item  \nameref{6}
    \item  \nameref{7}
    \item  \nameref{8}
    \item  \nameref{9}
    \item  \nameref{10}
    \item  \nameref{11}
    \item  \nameref{dimen_red}
    \item  \nameref{dimen_2}
    \item  \nameref{dimen_3}
    \item  \nameref{conclu}
\end{enumerate}

\section{Workflow in renal disease}\label{1}
After obtaining the sample by renal biopsy, there are three usual types of microscopy techniques that are applied to generate pathology images for diagnosis: light microscopy, immunofluorescence, and electron microscopy.

\section{Light microscopy (LM)}\label{2}

In light microscopy, the tissue sample obtained must first be fixed to eliminate metabolic activity. In a series of consecutive steps, non-structural contents such as water will be replaced with various substances. Finally, by adding a structure that provides rigidity, fine slices will be cut. On these slices, the staining protocols will be carried out and the dyes will adhere differentially to the cellular structures. Specifically for renal pathology, firstly, after obtaining the sample, it must be processed through the implementation of formaldehyde buffer and paraffin embedding. Subsequently, fine deparaffinized slices will be obtained and subjected to the following stains: periodic acid-silver methenamine, Masson's trichrome, periodic acid-Schiff, and hematoxylin and eosin\cite{zhang2020value}.

\subsection{Periodic acid-silver methenamine}
This technique is one of those used in nephropathology. The periodic acid of this stain oxidizes the carbohydrates present in the basement membranes of the sample, producing aldehydes. These aldehydes then reduce the silver to visible metallic silver\cite{bonsib2014atlas}.

\subsection{Masson's trichrome}
This technique combines, as the name indicates, three colors to differentiate structures. In this way, the structures can be separated into\cite{cohen1976masson}:

\begin{itemize}
    \item Light blue: basement membranes and mesangial matrix.
    \item Dark blue: collagen.
    \item Rust-colored: \begin{itemize}
        \item Finely granulated: cytoplasm of endothelial and mesangial cells.
        \item Coarsely granulated: cytoplasm of epithelial cells. 
    \end{itemize}
    \item Orange: proteins.
    \item Red: fibrin.
    \item Dark brown: nuclei. 
\end{itemize}

\subsection{Periodic acid-Schiff}
This technique also makes it possible to clearly distinguish the basement membranes, which stain bright pink, and the cell nuclei, which stain purple-blue\cite{bonsib2014atlas}.

\subsection{Hematoxylin and eosin}
This technique, the most widespread and common in medical pathology, stains cell nuclei purple-blue (hematoxylin) and the cytoplasm and basement membranes pink (eosin)\cite{bonsib2014atlas}.

\begin{figure}[h!]
\centerline{\includegraphics[scale=0.35]{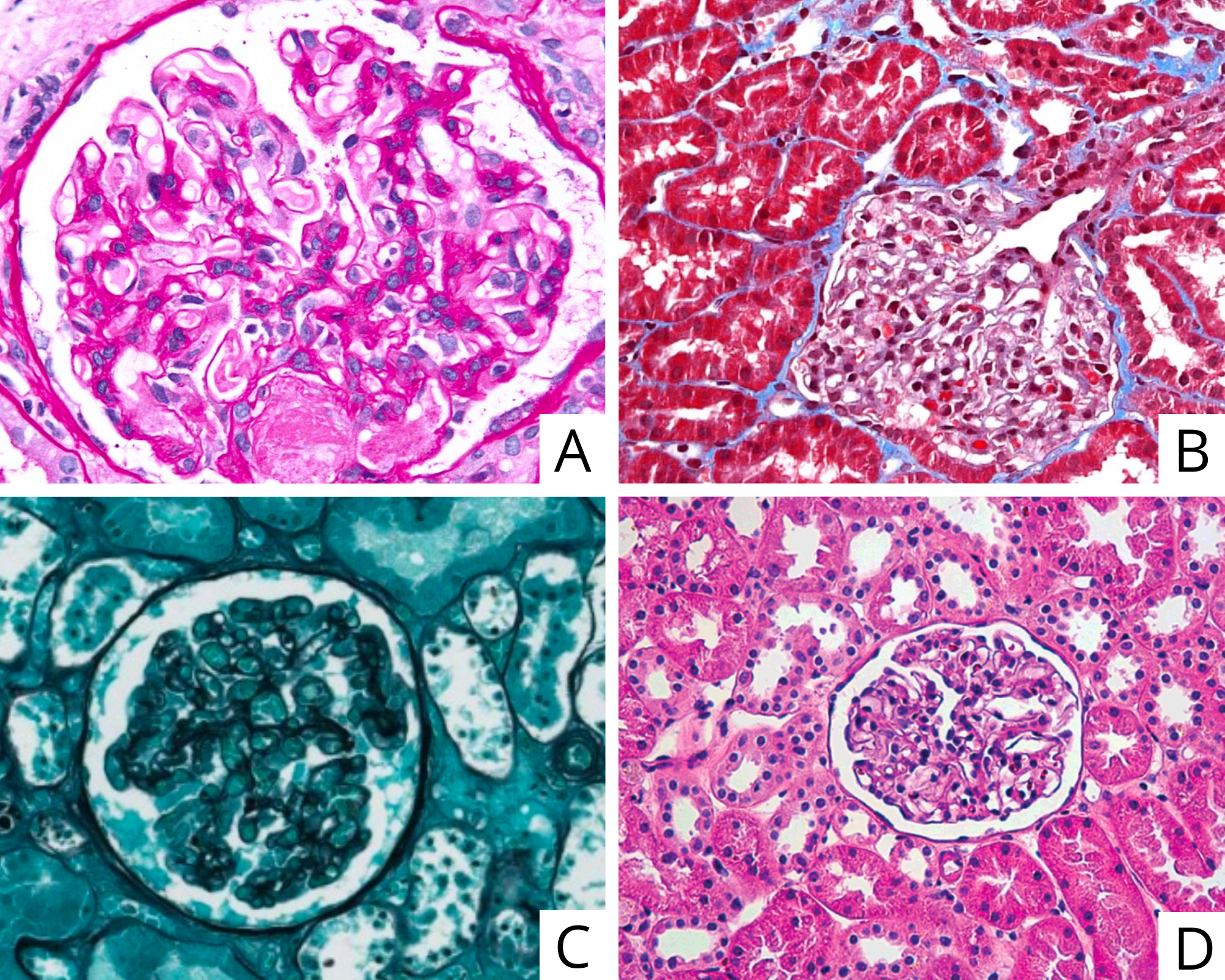}}
\caption{Examples of different stainings of a Bowman's capsule in kidney biopsies. A, renal cortex taken from a patient with acute thrombotic microangiopathy stained with Periodic acid-Schiff\cite{benz2010thrombotic}; B, normal renal cortex stained with Mason's trichrome\cite{farouk2019design}; C, normal renal cortex stained with periodic acid-silver methenamine\cite{farouk2019design}; D normal renal cortex stained with hematoxilin and eosin\cite{zhou2017silva}.}
\label{examples_patho}
\end{figure}

\section{Immunoflourescence (IF)}\label{3}

In immunofluorescence, unlike common histology, the stains are antibodies that adhere to target biomolecules (antigen) identifying a particular tissue portion. The antibodies contain a fluorophore, which after being incubated with the target, bind specifically to it. After incubation, the samples are observed under a fluorescence microscope, which excites the fluorophore with a laser. The fluorophore, after excitation, will return the energy taken from the laser in the form of light of a certain wavelength. Finally this light will be captured by a sensor in the microscope, processed and digitally built in the form of an image. In the specific case of immunofluorescence applied to nephropathology, we have two general approaches.

\subsection{Direct immunoflourescence (DIF)}
The most widespread and oldest is DIF, which is performed directly from a tissue sample prepared for this specific purpose: frozen tissue. It is important to note that in order to execute this approach, planning must be carried out in advance of the patient's biopsy and the necessary precautions must be taken to avoid the degradation of the sample during handling.

\subsection{Immunofluorescence on paraffin embedded biopsies (PIF)}
Many times it is not possible to have frozen tissue for various reasons. Firstly, because it could be not established at the time of the biopsy to carry out a diagnosis by immunofluorescence. Secondly, because the samples have been stored for a long time and are not suitable for the procedure. Or thirdly, simply because it is an interconsultation from another hospital where it is not possible to send a frozen sample by mail\cite{singh2016immunofluorescence}. In these cases, it is possible to resort to carrying out immunofluorescence from a paraffin block. It is important to note that paraffin induces protein crosslinking in the sample, which blocks the binding of antibodies to antigens. Therefore, the paraffin sample must be subjected to a process called antigen retrieval, in which proteolytic enzymes (trypsin, proteinase K or pronase E) or heat can be used to disentangle the protein network. Once the antigens have been exposed, the immunofluorescence protocol can be carried out\cite{zaqout2020immunofluorescence}.
Although this technique began as a backup for moments when frozen tissue was not available, something crucial was discovered about antigen retrieval. This process succeeded in unmasking antigens in some particular samples that were negative in DIF. That is, antigens that were not seen in the direct technique could only be observed through the previous step in paraffin\cite{messias2015paraffin}. In this way, PIF was established as a technique of additional diagnostic value, in addition to being a support for case consultations without frozen tissue samples\cite{singh2016immunofluorescence, messias2015paraffin}.

\subsection{Antibodies for renal pathology}
\begin{figure}[h!]
\centerline{\includegraphics[scale=0.35]{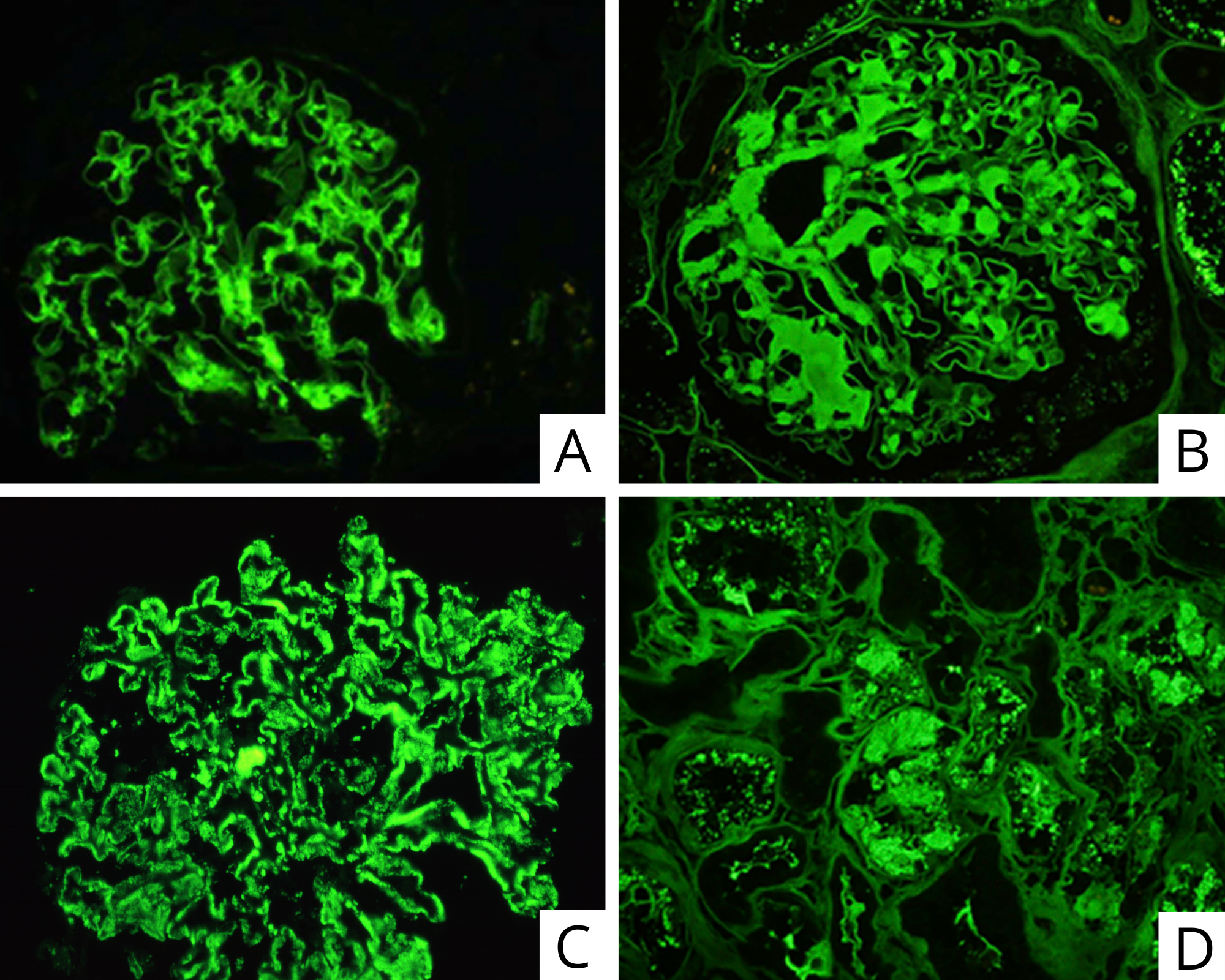}}
\caption{Examples of DIF and PIF images in kidney biopsies. A, PIF of renal cortex taken from a patient with IgA nephropathy, with antibodies against IgA\cite{nasr2018paraffin}; B, PIF of renal cortex taken from a patient with lambda-type AL amyloidosis, with antibodies against lambda\cite{nasr2018paraffin}; C,  DIF of renal cortex taken from a patient with membranous lupus nephritis, with antibodies against IgG\cite{fogo2021diagnostic}; D, PIF of renal cortex of a patient with kappa-type light chain proximal tubulopathy, with antibodies against kappa\cite{nasr2018paraffin}.}
\label{examples_immuno}
\end{figure}

The usual routine antibodies that are usually implemented in renal pathology for immunofluorescence are:
\begin{itemize}
    \item Heavy chains: IgA (Fig. \ref{examples_immuno}A), IgG (Fig. \ref{examples_immuno}C) and IgM.
    \item Light chains: Kappa (Fig. \ref{examples_immuno}D) and Lambda (Fig. \ref{examples_immuno}B).
    \item Complement: C3 and C1q.
    \item Cellular components: Fibrinogen and Albumin.
\end{itemize}

\section{Immunohistochemistry (IHC)}\label{4}
In IHC we use the same principle of specific binding between antibodies and antigens as in immunofluorescence. The difference is that in this procedure, the antibody is conjugated with an enzyme (for example, peroxidase) that catalyzes the color change of a substrate. In the specific case of peroxidase, the primary substrate is colorless and after the reaction, it turns brown\cite{shen2012role, al2011role}.

\begin{figure}[h!]
\centerline{\includegraphics[scale=0.35]{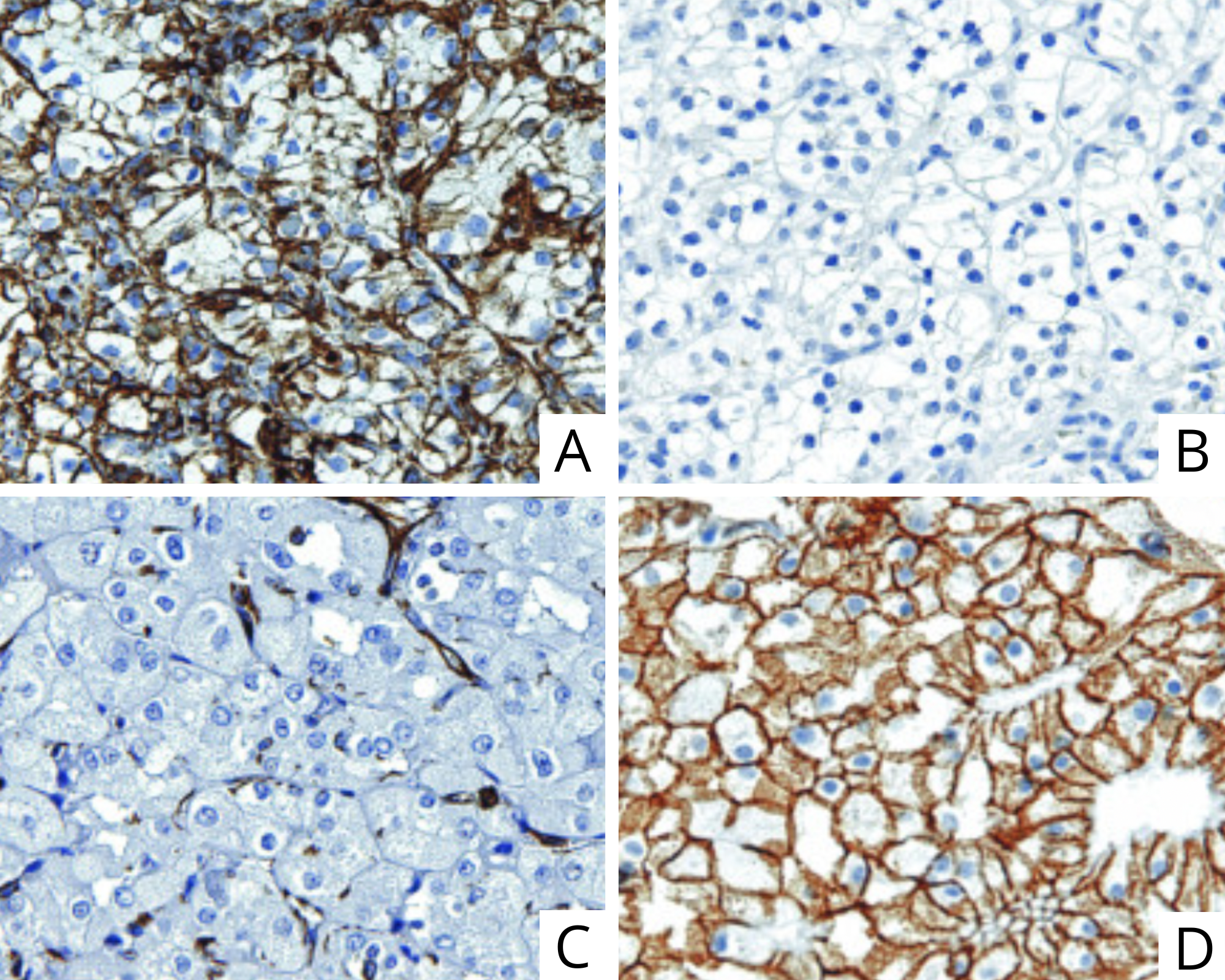}}
\caption{Examples of IHC in renal cell carcinoma (RCC) to separate clear cell subtype from chromophobe\cite{williams2009cd}. A, clear cell RCC with IHC against Vimentin; B clear cell RCC with IHC against CD9; C, chromophobe RCC with IHC against Vimentin; D, chromophobe RCC with IHC against CD9.}
\label{examples_IHC}
\end{figure}

In the specific case of kidney disease, this technique is used to identify tumor markers. Within the tumors found in nephrectomies we can find renal cell carcinomas (clear cell, papillary and chromophobes) and also oncocytomas. Among the markers usually studied with IHC, we find\cite{shen2012role, perna2007renal, al2011role}:

\begin{itemize}
    \item Cytokeratin: identification of mainly renal cell carcinoma. 
    \item Vimentin: identification of mainly renal cell carcinoma, clear cell and papillary (Fig. \ref{examples_IHC}).
    \item Transcription factors PAX8 and PAX9: identification of mainly renal cell carcinoma.
    \item Renal cell carcinoma marker: identification of mainly renal cell carcinoma.
    \item CD10: identification of mainly renal cell carcinoma (clear cell and papillary)
    \item CD9: identification of renal cell carcinoma, chromophobe (Fig. \ref{examples_IHC}).
    \item Cadherins E and kidney-specific: identification of mainly renal cell carcinoma (chromophobe ) and oncocytomas.
    \item Parvalbumin: identification of renal cell carcinoma (chromophobe) and oncocytomas.
    \item Claudins 7 and 8: identification of renal cell carcinoma (chromophobe) and oncocytomas.
    \item CD117: identification of renal cell carcinoma (chromophobe) and oncocytomas.
    \item Alpha-Methylacyl Coenzyme A Racemase (AMCAR): identification of renal cell carcinoma (papillary).
    \item Transcription factors TFE3 and TFEB: mainly involved in pediatric renal cell carcinomas. TFE3 associated with a translocation that involves Xp11.2.
    \item Cathepsin-K: identification of renal cell carcinoma products of translocations. 
    \item Uroplakin III: identification of renal cell carcinoma of high grade. 
    \item p63: identification of renal cell carcinoma of high grade. 
    \item Thrombomodulin: identification of renal cell carcinoma of high grade. 
    \item GATA3: identification of renal cell carcinoma of high grade. 
\end{itemize}

\section{Electron microscopy (EM)}\label{5}
In renal pathology, EM is a necessary partner to diagnose a large number of cases, where the resolution of optical microscopy is not sufficient. In different applications of nephropathology, percentages of cases have been reported where EM adds relevant information for diagnosis (Table \ref{EM_importance}). Of those case reports where EM was necessary, the mean was 49\%. This means that from the previous publications, which covered from 1997 to 2020, in 50\% of the cases of nephropathology it was necessary to resort to EM to reach the diagnosis.
Within renal pathologies, there is a group of them\cite{mokhtar2011role} that require the resolution of electron microscopy in order to arrive at a diagnosis (Table \ref{Conditions_EM}) For example, in cases such as focal segmental glomerulosclerosis, being able to determine the degree and extent of foot process effacement by EM is key to distinguishing between a primary or secondary condition ((Table \ref{Conditions_EM}). Additionally, of the conditions listed, there are 6 that have one or more associated genes. These pathologies (Fabry, Nail-Patella, trans-retinein amyloidosis, fibronectin glomerulopathy, Alport's and thin basement membrane) also have confirmation by genetic sequencing that can help determine the diagnosis without necessarily having to resort to EM. However, to stage the progress of those pathologies, it is necessary to have the microscopic image.

\subsection{Procedure}

\begin{figure}[h!]
\centerline{\includegraphics[scale=0.35]{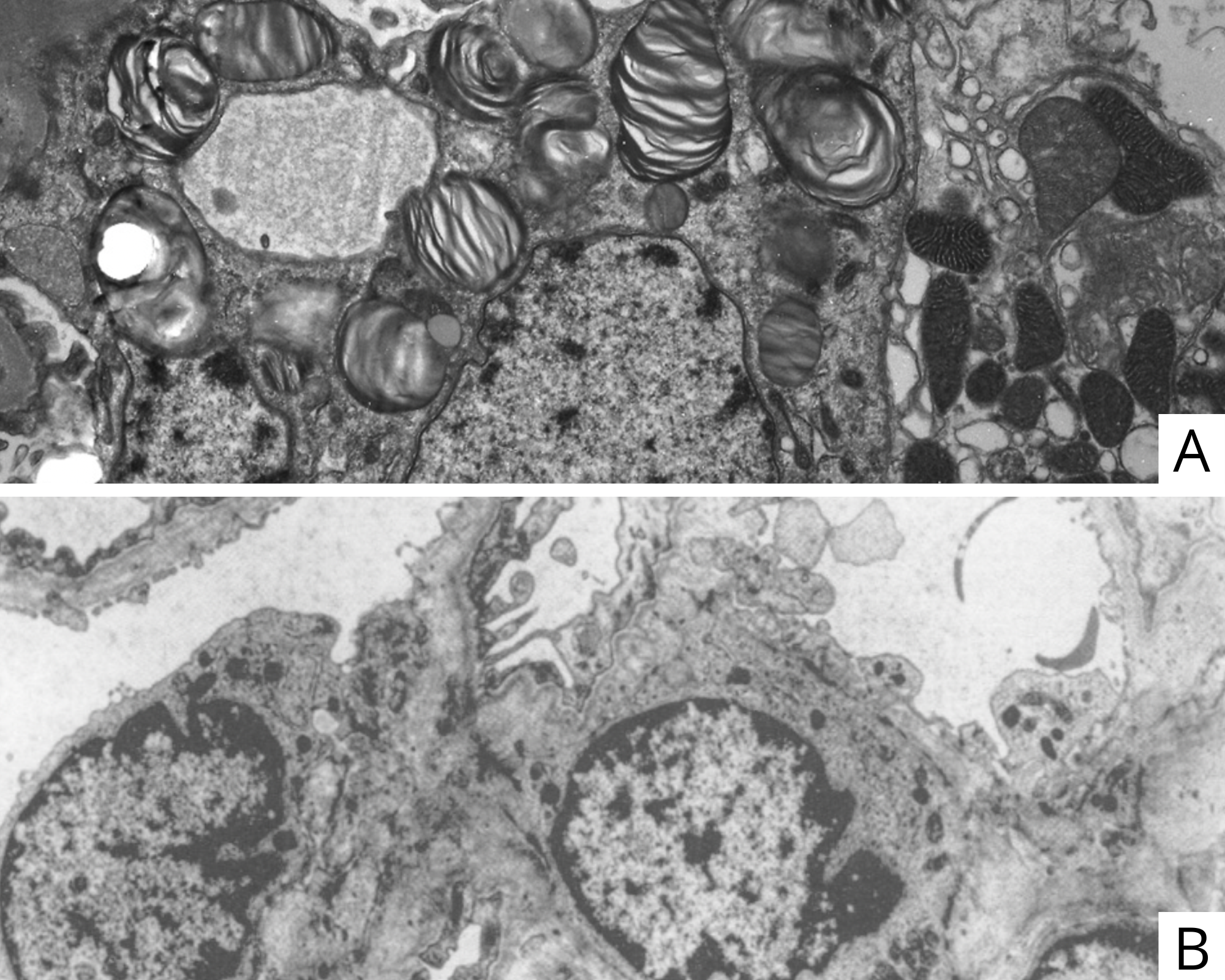}}
\caption{Examples of electron microscopy (EM) images of renal biopsies. A, renal biopsy observed with EM showing zebra bodies of a Fabry disease patient\cite{sugarman2018atypical}; B, renal biopsy observed with EM showing the enlarged and split glomerular basement membrane (GBM) of an Alport's syndrome patient\cite{flinter1997alport}.}
\label{examples_EM}
\end{figure}

The general protocol for processing kidney biopsies for EM begins with the division of the specimen into 0.5 mm portions and then fixation with 3\% glutaraldehyde and osmium tetroxide. Subsequently, the sample must be dehydrated and embedded with SPI-EPON 812. Once the sample is embedded,  slices must be made with a glass blade microtome. Finally, when the slices are produced, they must be stained with uranyl acetate and lead citrate, which, due to their density, will absorb the electrons to a greater extent.
According to the type of cells and their content, the citrates will be absorbed differentially, resulting in zones with different concentration (Fig. \ref{examples_EM}).
Among these zones, we are going to have ones with more citrate that will absorb more electrons (darker) and others with less citrate (lighter)\cite{howell2021electron, johannessen1973rapid, zhang2020value}.

\begin{table*}[]
\caption{List of publications with the percentage of cases where it was necessary to resort to electron microscopy (EM) for diagnosis in nephropathology.}
\begin{center}
    
\begin{tabular}{llllll}
\hline
\textbf{Application}                  & \textbf{\% of cases} & \textbf{Total cases} & \textbf{Location}    & \textbf{Date} & \textbf{Reference}                                       \\ \hline
Kidney transplant                     & 50                   & N/A                  & Louisiana, USA       & 1997          & \cite{haas1997reevaluation, silva1988electron, herrera1997role} \\
Chronic renal allograft rejection     & 50                   & 91                   & Szeged, Hungary      & 2001          & \cite{ivanyi2001value}                                          \\
General nephropathy                   & 72                   & 133                  & Turku, Finland       & 2003          & \cite{collan2005value}                                          \\
Glomerular disease                    & 38                   & 52                   & Tunis, Tunisia       & 2006          & \cite{darouich2010value}                                        \\
Glomerular disease                    & 20                   & N/A                  & Minnesota, USA & 2006          & \cite{sethi2016mayo}                                            \\
Kidney disease after liver transplant & 80                   & 81                   & New York, USA        & 2010          & \cite{kim2010variable}                                          \\
General nephropathy                   & 39                   & 273                  & Jeddah, Saudi Arabia & 2011          & \cite{mokhtar2011role}                                          \\
General nephropathy                   & 31                   & N/A                  & Kathmandu, Nepal     & 2013          & \cite{pant2013role}                                             \\
Pediatric nephropathy                 & 68                   & 855                  & Peking, China        & 2020          & \cite{zhang2020value}                                           \\ \hline
\end{tabular}
\end{center}
\label{EM_importance}
\end{table*}

\begin{table*}[]
\caption{List of renal conditions that require electron microscopy (EM) to be diagnosed. GBM, glomerular basement membrane; BM, basement membrane; FP, foot process; TM, tubules membrane.  }
\begin{tabular}{llll}
\hline
\textbf{Condition}                           & \textbf{Gene}  & \textbf{Microscopic observation}                                                                             & \textbf{Reference}              \\ \hline
Fabry                                        & GLA            & Zebra bodies / Myelin figures                                                                    & \cite{de2017ultrastructural}           \\
Nail-Patella                                 & LMX1B          & Zebra bodies / Myelin figures                                                                    & \cite{edwards2015novel, lei2020myelin} \\
Trans retinein amyloidosis                   & TTR            & Amyloid deposits                                                                                 & \cite{nishi2008new}                    \\
Light chain amyloidosis                      & NO             & Amyloid deposits                                                                                 & \cite{nishi2008new}                    \\
Heavy chain amyloidosis                      & NO             & Amyloid deposits                                                                                 & \cite{nishi2008new}                    \\
Serum amyloid A amyloidosis                  & NO             & Amyloid deposits                                                                                 & \cite{nishi2008new}                    \\
Fibrillary glomerulonephritis                & NO             & Non-branching fibrillary deposits                                                                & \cite{lusco2015ajkd}                   \\
Fibronectin glomerulopathy                   & FN1            & Granular and fibrillary deposits                                                                 & \cite{lusco2017ajkd}                   \\
Collagenofibrotic glomerulopathy             & NO             & Deposits of collagen bundles                                                                     & \cite{kurien2015collagenofibrotic}     \\
Alports                                      & COL4A3, COL4A4 & Basket-weave pattern                                                                             & \cite{fogo2016ajkd, kashtan2018alport} \\
Thin basement membrane                       & COL4A3, COL4A4 & Diffuse thinning of the GBM                                            & \cite{foster2005pathology}             \\
Diabetic nephropathy                         & NO             & Diffuse thickening of GBM, nonatrophic TM thickened      & \cite{najafian2015ajkd}                \\
Dense deposit disease                        & NO             & Dense transformation of the BM lamina densa                                       & \cite{fogo2015ajkd}                    \\
Minimal change disease                       & NO             & Foot process effacement.                                                                         & \cite{fogo2015ajkd1}                   \\
Focal segmental glomerulosclerosis primary   & NO             & Extensive foot process effacement                                                                & \cite{fogo2015ajkd2}                   \\
Focal segmental glomerulosclerosis secondary & NO             & Imperceptible FP effacement                                                            & \cite{fogo2015ajkd2}                   \\
Podocyte infolding glomerulopathy            & NO             & Intramembranous microspherules, Invagination of FP into GBM & \cite{ting2021podocyte}                \\
Transplant glomerulopathy                    & NO             & GBM reduplication                                                       & \cite{filippone2018transplant}         \\ \hline
\end{tabular}
\label{Conditions_EM}
\end{table*}

\section{Hyperspectral imagery (HI)}\label{6}

\begin{figure}[h!]
\centerline{\includegraphics[scale=0.5]{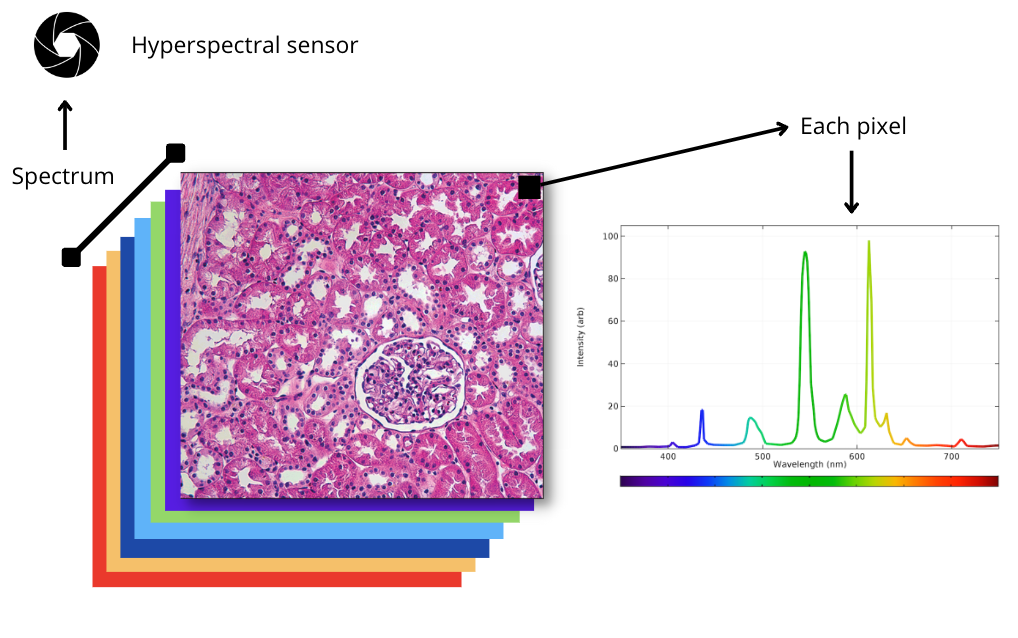}}
\caption{Representation of hyperspectral imagery on kidney tissue. Each color represents a wavelenght and each pixel of the image contains a full spectrum.}
\label{examples_hypers}
\end{figure}

This method consists of capturing multiple planes of information from the tissue sample. The human eye is capable of viewing objects in a wavelength range of 380 to 740 nanometers (nm). Hyperspectral imagery can extend this range to 2500 nm\cite{ortega2020hyperspectral}. In this way, the sensor will take multiple snapshots of the same sample at different wavelengths (for example, 400, 450, 500 and 600 nm). This will generate several arrays for each wavelength (Fig. \ref{examples_hypers}, representation of the left). Consequently, each pixel will have information on all the wavelength planes that were recorded. So each pixel will have its own spectrum graph also called spectral signature (Fig. \ref{examples_hypers}, graph on the right)). The spectral signature is simply the representation of the intensity and wavelength recorded in each capture\cite{ortega2020hyperspectral,tu2021deep }.
This method has been applied in the field of pathology, since healthy tissue and pathological tissue present different spectral signatures\cite{callico2017image, tu2021deep}. This difference is the key to detecting pathological tissue, where a bright field image is not capable of providing information. 
With regard to applications in renal pathology, thanks to hyperspectral imaging it has been possible to separate and distinguish two membranous nephropathies. In these two highly similar pathologies, one is caused by a virus (hepatitis B virus) and the other is idiopathic\cite{tu2021deep}.

\subsection{Projection transformation}
Since the data of a hyperspectral image is multidimensional, it is advisable to be able to use dimensionality reduction techniques that also help reduce noise. Among the most used we can find the principal component analysis (PCA) or Fisher's linear discriminant analysis (LDA). These techniques have their negative counterpart, which lies in the need for the data to have a Gaussian distribution. Therefore, there is another technique called local Fisher's discriminant analysis(LFDA)\cite{tu2021deep}, developed after the previous ones and that does not need a Gaussian distribution to operate.

\section{Renal computational pathology}\label{7}

In the previous sections we have seen very briefly the different techniques of histology and image acquisition to work in renal pathology. In the next sections, we will inquire about the state of research in computational pathology applied to nephrology.

\subsection{Common tasks for renal pathology}

After a review of the literature, we can agree with the position of the excellent article developed by Huo \emph{et al} in 2021\cite{huo2021ai}. The main tasks in which computer vision algorithms are used in renal pathology are:

\subsubsection{Classification}
when we refer to classification in computer vision, we mean to assign a label to an image. In this way, an image enters the algorithm and the algorithm returns a label with a probability (Fig. \ref{examples_renal_AI}). Given that in pathology,  whole slide images (WSIs) have a considerable size, it is customary to partition them into patches. All patches that correspond to a WSI will then be entered into the algorithm. The algorithm will set one label per patch. Then, with all the labels, a heatmap can be made as a visual aid on top of the WSI. A general label of the WSI can also be established (for example, if we identify patches with cancer, the entire WSI will correspond to a cancer patient)\cite{goodfellow2016deep, huo2021ai}.

\subsubsection{Detection}
the main difference between classification and detection is that in classification we get only one label in the entire patch. It doesn't matter if inside this patch there are also other things of interest. However, in detection, in addition to being able to determine several regions of interest in the same area, we can also determine the position. This is especially important in renal pathology, to identify functional units (such as the glomerulus) or to identify anatomical parts (arteries, capillaries or membranes). Within the most used techniques for detection, we find the bounding box(Fig. \ref{examples_renal_AI}). This technique consists of the creation of an artificial rectangle that identifies the object of interest (for example, the glomerulus\cite{heckenauer2020real}) in an image. In this way, the input is the same as in the classification: an image. However the output is different: in detection the output is the image plus the rectangle where the region of interest is\cite{goodfellow2016deep, huo2021ai}.

\subsubsection{Segmentation}

when we allude to segmentation, we are alluding to the process of assigning each pixel of an image a particular label. Thanks to knowing exactly how many pixels we have with a certain label, we can perform quantification tasks within an image. For example, in certain neurological diseases it is possible to measure the area within a certain image that is affected and thus determine the diagnosis\cite{jose2014brain}. In pathology, segmentation is basically used to separate healthy tissue from pathological tissue or to identify functional units (such as the glomerulus)\cite{goodfellow2016deep}. The segmentation input is an image and the output is a mask(Fig. \ref{examples_renal_AI}). The mask is superimposable on the image, so it indicates the pixels of interestcite\cite{goodfellow2016deep, huo2021ai}.

\begin{figure}[h!]
\centerline{\includegraphics[scale=0.48]{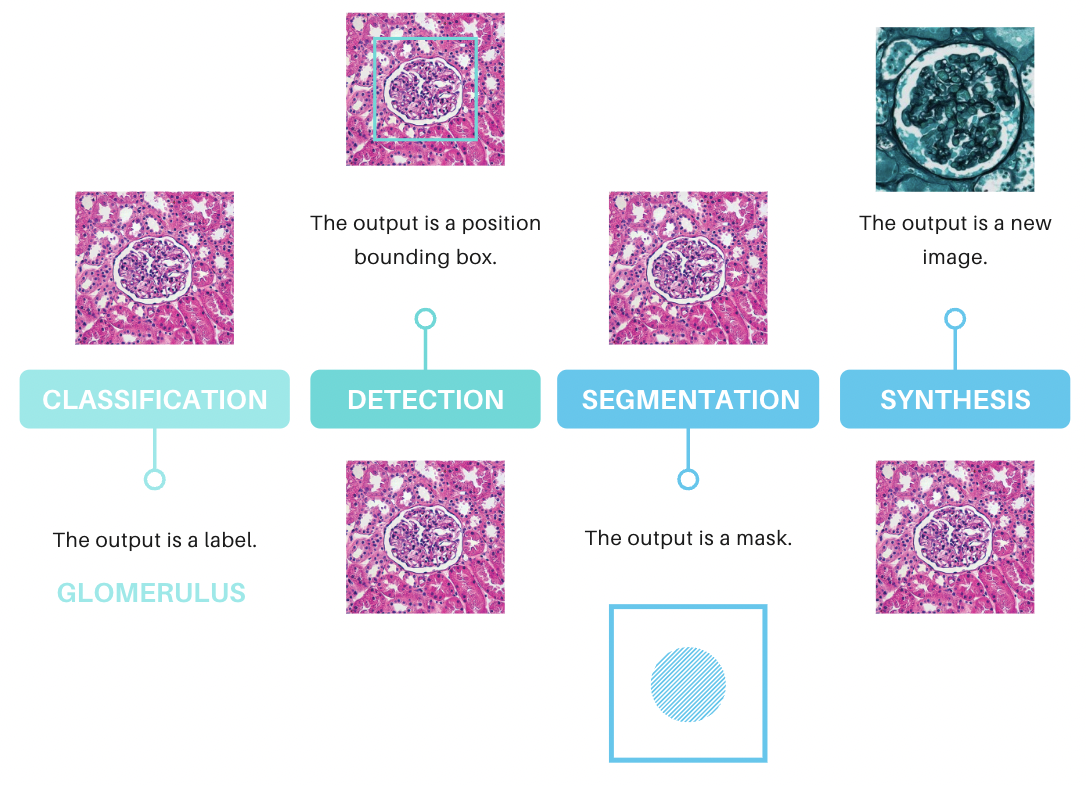}}
\caption{Representation of common tasks in computer vision applied to renal pathology.}
\label{examples_renal_AI}
\end{figure}

\subsubsection{Synthesis}\label{Synthesis}

unlike the other three approaches, image synthesis produces new images. To produce these new images, we must first train on the model images. The generator model will learn from the distribution of the training images and will generate new images very similar to the training ones(Fig. \ref{examples_renal_AI}). However, another model, called a discriminator, will try to discover which are the artificial images within a batch of artificial and real images\cite{liu2021review}. This pairing between generator and discriminator will serve to produce in each cycle, artificial images more and more similar to the real ones. In renal pathology, synthesis is used to produce artificial stains. In this way, the network is trained so that it can generate a stain from an image with another stain (for example, H\&E to PAS)\cite{goodfellow2016deep, huo2021ai}.

\section{Frequent use cases in renal computational pathology}\label{8}

After our review of the literature, we can structure and detail some frequent uses of algorithms in renal pathology.

\subsection{Glomeruli identification}
Of the many research works and prototypes analyzed, there is a large group of them that focus on the identification of glomeruli\cite{heckenauer2020real,
lo2018glomerulus,
kannan2019segmentation,
 jiang2021deep,
 hermsen2019deep,
 hara2022evaluating,
 de2018automatic,
 marsh2018deep,
jayapandian2021development,
temerinac2017detection,
hao2020classification,
 zheng2021deep,
altini2020deep,
salvi2021automated,
marsh2021development,
patterson2021autofluorescence,
wilbur2020using}. This identification can be developed by segmentation\cite{ kannan2019segmentation,jiang2021deep,hermsen2019deep,hara2022evaluating,de2018automatic, jayapandian2021development,salvi2021automated,patterson2021autofluorescence} or detection\cite{heckenauer2020real,lo2018glomerulus,temerinac2017detection,hao2020classification, zheng2021deep,altini2020deep,marsh2021development,wilbur2020using}. After the recognition of this functional unit, it is possible to proceed to separate pathological from healthy. Also, after this separation, the number of healthy and pathological glomeruli can be quantified, to stage the patient. Of the stains used in these prototypes, we find the most frequent, H\&E\cite{heckenauer2020real,lo2018glomerulus ,jiang2021deep,marsh2018deep,jayapandian2021development,temerinac2017detection,marsh2021development,wilbur2020using}, and also the others typical of renal pathology: PAS\cite{lo2018glomerulus, jiang2021deep,hermsen2019deep,hara2022evaluating, de2018automatic,jayapandian2021development,temerinac2017detection,zheng2021deep,salvi2021automated,patterson2021autofluorescence,wilbur2020using}, PAM\cite{jiang2021deep, hao2020classification}, Masson\cite{kannan2019segmentation,jiang2021deep,jayapandian2021development, wilbur2020using}, Silver\cite{jayapandian2021development, wilbur2020using} and IF\cite{patterson2021autofluorescence}. The mean number of patients in these studies was 182, while the study with the largest number had 1281\cite{hao2020classification} and the one with the least number 10\cite{heckenauer2020real}. These prototypes used fliping, rotation, cropping, scaling, contrast and brightness changes, noise adding, transposition and HUE color space transition as data augmentation techniques. Among the most frequent applications of the prototypes we find renal transplantation\cite{heckenauer2020real,hermsen2019deep,de2018automatic,altini2020deep} and the identification of glomerulosclerosis\cite{marsh2018deep, salvi2021automated,marsh2021development}.

\subsection{Lesion identification}

Following a similar reasoning to the one exposed in the previous section, some prototypes focused on the detection of lesions in renal tissue\cite{zeng2020identification,kers2022deep,pan2021multi,barros2022podnet}. The most frequent pathology found in this group of studies was IgA nephropathy\cite{zeng2020identification, pan2021multi}. Most of the prototypes used PAS-stained\cite{zeng2020identification, kers2022deep, barros2022podnet} images and only one used IF\cite{pan2021multi}.

\subsection{Cancer cells identification}
Like most of the research papers in oncological digital pathology, a group of the listed papers focused specifically on the identification of tumor cells in renal tissue. Of this group, all were performed in renal cell carcinoma with H\&E staining\cite{ tabibu2019pan, fenstermaker2020development, abdeltawab2022deep, zhu2021development}. Two of these works were carried out with images extracted directly from The Cancer Genome Atlas (TCGA)\cite{tabibu2019pan, fenstermaker2020development}.

\subsection{Condition identification}

Part of the research work was also focused on the identification of specific medical conditions. The techniques used were EM\cite{hacking2021deep}, PAS/PAM\cite{uchino2020classification} and IF\cite{kitamura2020deep}. One article in particular\cite{hacking2021deep} focused on a group of pathologies such as amyloidosis, diabetic glomerulosclerosis, membranous nephropathy, membrano-proliferative glomerulonephritis and thin basement membrane disease.

\subsection{Stain generation}
As we mentioned before (Section \ref{Synthesis}), the possibility of generating virtual stains is of vital importance to extend datasets. Two research papers focused on this process. The first of them generating images of Masson trichrome, Jones and PAS from a training with H\&E images for the diagnosis of non-neoplastic kidney diseases\cite{de2021deep}. Second, IF images are generated from brightfield PAS images, computationally identifying cellular substructures such as cell nuclei, collagen, fibrosis, distal tubules, proximal tubules, endothelial cells and leukocytes\cite{jen2021silico}.

\subsection{Individual entities identification}

With respect to the identification of individual pathological or functional entities, we have several articles of this type. Among them we find the identification of fibrosis\cite{zheng2021deep}, macrophages and lymphocytes\cite{hermsen2021quantitative}, deposits\cite{tu2021deep}, peritubular capillaries\cite{kim2019fully}, restricted vascularity\cite{huang2017novel} and deposits in glomeruli\cite{ligabue2020evaluation}. Other articles instead resort to the segmentation of structures to produce a more accurate pixel map. In this group we find podocyte segmentation\cite{zimmermann2021deep} and tubule segmentation\cite{yi2022deep}.
Additionally, most of the prototypes that worked on the identification of deposits used some type of immunological technique (IHC or IF)\cite{hermsen2021quantitative, yi2022deep, kim2019fully, huang2017novel}. Also, the only peer-reviewed publication that includes the use of hyperspectral imaging was performed on deposits identification for membranous nephropathy screening\cite{tu2021deep}.

\section{Image aligning} \label{9}
\begin{figure}[h!]
\centerline{\includegraphics[scale=0.45]{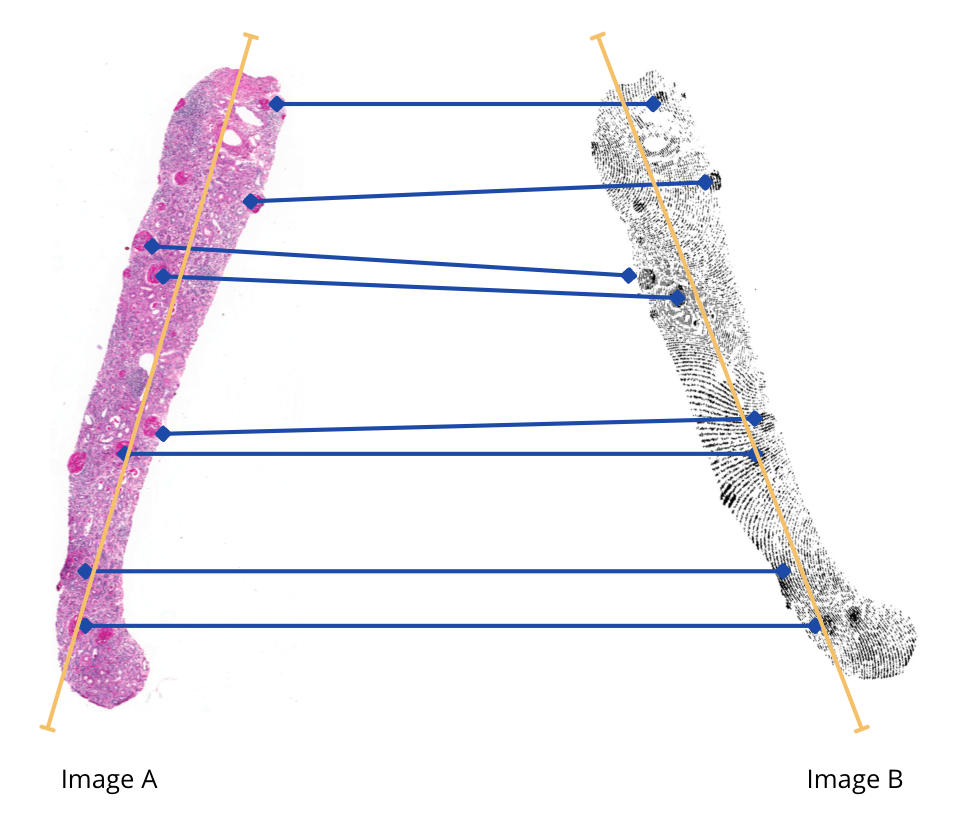}}
\caption{Representation of the image registration process proposed by \cite{wilbur2020using}. Image A, is the first image with one stain. Image B, is the second image with a different stain that come from the same surgical piece. Each blue point represents a coordinate and the blue lines concatenate corresponding points. Yellow lines represent the axis of each image. }
\label{examples_registration}
\end{figure}

One of the biggest challenges in optimizing image labeling work is aligning samples of the same tissue. When performing multiple stains on successive slices of the same surgical specimen, being able to enlarge the image in the same position on all stains is key. Since the images that come from the same surgical piece represent different planes of it, a degree of similarity is expected. However, given the change in plane and the tissue's own deformation, the differences can be widened. For example, if we are observing a glomerulus in H\&E and we want to see if the basement membrane is modified, we can observe that in the PAS image of the same gromerulus. To do this, we would have to locate the gromerulus in the PAS image, which will surely be in another tissue cut plane.
There is a technique specifically developed for renal tissue images that allows aligning images of the same surgical piece subjected to successive staining. This technique is called image registration\cite{wilbur2020using}. It consists of grouping images that come from the same origin (surgical piece) and spatially distributing affine transform matrices throughout the extension of each image. These matrices are basically recognition points (tissue characteristics) that are repeated in all the images (Fig. \ref{examples_registration}). When they have all been identified, they are concatenated in all the images\cite{wilbur2020using}.
The recognition points identified above are coordinates. Each coordinate has a unique code. This allows that when an important point is identified in one image (glomerulus in H\&E), the same position can be quickly found in another image (glomerulus in PAS).

\section{Data protection in renal computational pathology}\label{10} 
\subsection{Informed consent}

Regarding data protection, many articles submitted the justification for the study to the ethics committee of the institution that provided the data (following the Helsinki declaration). The committee itself authorized the execution of the study. It also established the waiver of informed consent justifying that the study was retrospective and does not involve any further tests or treatments of the participants (the samples were already present in the biobank)\cite{zeng2020identification, zheng2021deep, altini2020deep}.
Other articles used databases that had already submitted their donors to informed consent for the use of their data. In this way, the authors, as long as they are compliant with the conditions of use of the information and the protection regulations, should not resort to new consents\cite{tabibu2019pan, fenstermaker2020development, jayapandian2021development}. 
Some works did not mention in the text of the article or in any appendix if they submitted the study to an ethics committee or if they received an informed consent waiver to proceed with the protocols. It is possible that the information is available in another link or section, which is not directly visible to users\cite{de2018automatic, marsh2018deep}.
Other studies mention directly in the final statement that they have followed the ethical standards of the Helsinki declaration. However, they do not mention having submitted the protocol to validation by the ethics committee or having had any type of content review by any authority of the partner institution\cite{hacking2021deep}.

\section{Interpretability}\label{11}

Interpretability of an algorithm can be defined as a set of protocols, techniques or properties that make the output generation of that algorithm understandable by humans. When we refer to understandable, we mean that it can be inferred, for example, what characteristics of an image were those that defined its belonging to a category (label) in a classification\cite{graziani2021interpretability}.

\subsection{Dimensions of interpretability}

It has been proposed that three dimensions could be considered when attempting to interpret the output of an algorithm\cite{graziani2021interpretability, montavon2018methods}: 

\begin{itemize}
    \item The first dimension focuses on data. Try to apply techniques to understand which of the dimensions of our data are the most important to solve the task at hand. As we mentioned before (Section \ref{dimen_red}), the reduction of dimensions and the identification of variables that are correlated can help considerably to select the best features for our model. Having an insight into what features are involved in the task helps us partially understand how a decision is reached.
    \item The second dimension has to do with prediction. Specifically, how a certain input resulted in a certain output in the algorithm. For example, in an image what angles, silhouettes, shapes or colors are necessary for that image to be classified with a particular label. Examples of this dimension could be CAM (Section \ref{CAM}), GRAD-CAM (Section \ref{GRAD_CAM}), ablation GRAD-CAM (Section \ref{ablation_CAM}) and score CAM (Section \ref{score_CAM}).
    \item The third dimension has to do directly with the model. What parts of the model interact with each other and with what force so that the input that enters is processed to a respective output. In the specific case of neural networks, how the layers interact with each other and with what weight the information from a previous layer passes to a later one. Examples of this dimension could be considered activation maximization (Section \ref{acti_max}), feature visualization (Section \ref{featu_vis}) and network dissection (Section \ref{net_diss}).
  
\end{itemize}

\section{First dimension: data level interpretability}\label{dimen_red}
When we work with pathology images (WSI), we subject our hardware to a large impact on processing time and number of active units. Therefore, reducing the dimensions of our input data is key to:

\begin{itemize}
    \item reduce physical storage capacity
    \item reduce processing time
    \item increase response speed
\end{itemize}

When we talk about reducing the dimensions of our data, we are specifically referring to isolating the key information for the task we need to develop. That is, all that part of the information that would not contribute to the task for which the algorithm was built (noise), should disappear. This separation seems simple in words, but it is much more complex in practice\cite{cascianelli2018dimensionality}. We have several strategies to reduce the dimensions of our data, and we will detail them below.

\subsection{Principal component analysis (PCA)}

To perform this technique on our features, we must first project the values of those features (for each sample of our dataset) in a space of lower dimension. The type of projection must ensure that the variance of the data is the largest. Having the projected data, we will proceed to search for the main components. These components are vectors that best fit the data. Once we find these small vectors, they will be the components that will represent the thousands of values that previously belonged to these variables\cite{cascianelli2018dimensionality, bishop2006pattern}.

\subsection{Gaussian random projection (GRP)}
We can also resort to GRP, which simply consists of projecting our features in a space of reduced dimensions. To do this we simply multiply our matrix (features, n x m) by a smaller dimension matrix of random values (m x p) and a scalar (1/square root p). This scalar will help absorb the impact of the pairwise distance resulting from the dimension reduction\cite{cascianelli2018dimensionality}.

\subsection{High correlation feature elimination}
This method is based on the precept that groups of variables that have a high correlation among them can be represented by only one of the group. In this way, the correlation of each feature with each of its peers must be calculated. If the correlation is greater than a threshold, then one can be kept and the other eliminated. The higher the threshold, the less number of features will be discarded. Therefore, it is important to always perform a finetuning with different thresholds to find the optimal one\cite{cascianelli2018dimensionality}.

\subsection{Neural compression}
This procedure consists of using the encoder to generate embedded vectors. The encoder is fed with the images whose dimensions we want to reduce. The result will be a vector of reduced dimensions for each image. For WSI, there is a protocol by which the image is divided into patches. Then each patch is converted to a vector. When all the patches have been converted, the image is reassembled with the vectors preserving the location of each one\cite{tellez2019neural, cossio_2022}.

\subsection{t-distributed stochastic neighbor embedding (t-SNE)}
t-SNE is a very useful technique for reducing dimensions of multidimensional data. This technique uses the abstract representations (feature vectors) of the last layer of a neural network (before running softmax). Since these vectors are highly multidimensional, this technique reduces the dimensions in a non-linear way. After reduction, it produces a representation of the vectors according to their similarity. This can help us to know, in a multiclass classification, which are the most important features of each class and how similar they are with respect to other classes.

\subsection{Applications in renal computational pathology}
Of the dimensionality reduction techniques, we find some applications in digital pathology. One group managed to locate the most important features for the classification of different types of nervous system neoplasms, using images similar to those of renal pathology. Moreover, with the t-SNE they managed to represent them visualizing the different types of tumors by similarity\cite{faust2018visualizing}. Image compression was very regularly used as a method for dimensionality reduction of pathology images, in breast cancer\cite{tellez2019neural} and lung cancer\cite{aswolinskiy2021neural}. H\&E images were used in the same fashion as many renal pathology studies. PCA was also used as a color separator in patches of glomeruli, eliminating noise and isolating structural components more effectively than the original image\cite{sarder2016automated}. Regarding the correlation of features, a variation of the method explained above was used to select features correlated with the pathological class. In this way, the features with a correlation coefficient greater than a threshold for the pathological class were selected\cite{shen2022deep}.

\section{Second dimension: classification level interpretability}\label{dimen_2}

\subsection{Class Activation Mapping (CAM)}\label{CAM}

The CAM technique uses the feature maps of the last layer of the neural network. To generate a single usable result, all feature maps are linearly combined using the weights learned from the last layer as a guide. In this way, each feature map is multiplied by its weight and then the linear combination of all of them is produced. An important disadvantage of this technique is the inability to operate with fully connected layers in the last layer. Therefore, the networks that have this characteristic will have to transform their last layer to convolutional and retrain\cite{zhou2016learning}.

\subsection{Gradient-weighted Class Activation Mapping (Grad-CAM)}\label{GRAD_CAM}

This technique was built using some resources of the previous one. Gradients backpropagating from the output are computed for each class. The weights are calculated from the gradient, using global average pooling. These weights represent, like the previous technique, the importance of each feature map. However, after linearly combining these feature maps with their weights, ReLU is applied. This is the main difference. When applying ReLU, the features that only have a positive influence on the interest class are selected\cite{selvaraju2016grad, selvaraju2017grad}.

\subsection{Ablation CAM}\label{ablation_CAM}

Unlike the previous technique, in this one the calculation of the gradient is not made use of. Instead, a process called ablation is employed. To do this, the values of each feature map are set to zero per iteration. The drop in activation score is calculated for the given class, for each of the feature maps. The magnitude of the drop represents the relative importance of each feature map. Finally, following a protocol similar to the previous one, all the calculated importances are combined and ReLU is applied to prioritize only those with a positive effect\cite{ramaswamy2020ablation}.

\subsection{Score CAM}\label{score_CAM}

This technique, like the last one, gets rid of the calculation of the gradient. In this case, as a first step, they generate the feature maps of one of the final layers. As a second step, with these maps, masks are built and applied to the original input. Each new image (mask on input) is passed through the network again. Each pass will generate a score. Finally, all the generated feature maps will be linearly combined (first step) using the scores as weights (second step). ReLU is used at the end of the combination, to take only the positive influence\cite{wang2020score}.

\subsection{Applications in renal computational pathology}

Specifically concerning renal pathology, the most widely used interpretability techniques have been CAM and GradCAM\cite{zheng2021deep, weis2022assessment}. Regarding the first, it has been used both at the global level of WSI\cite{zheng2021deep} and at the patch level, with a focus on the gromerulus. When multiple pathologies are classified at glomerular level, the focus of the neural network can be seen in the patch. For example, in patients with amyloidosis, the network can be seen focusing on an area of the glomerulus where amyloid deposits are placed\cite{weis2022assessment}.
Regarding the technique score CAM, we found only one study that compared its use with respect to GradCAM in the evaluation of renal tissue in patients with immunoglobulin A nephropathy\cite{sato2021evaluation}. We can see in the images of the heatmaps, slight differences between one technique and another for the same image and the same class.
Regarding dark field images, GradCAM has been applied as well. It has been found that the technique is equally useful and that it is able to point out the regions of importance in the classification, even in digitally blurred images\cite{pan2021multi}.

\section{Third dimension: model level interpretability}\label{dimen_3}

\subsubsection{Activation maximization}\label{acti_max}
This method aims to find the parts of a certain input (for example, parts of an image) that the network considers to be the most important in order to assign a certain output to that input. In this way, once the model has been trained, this method finds all the features that make up a class. For example, if we train our model with numbers from MNIST, we can generate a list of representations that characterize the label '8'. This list would contain upper and lower curves that would filter out images resembling a number 8\cite{erhan2009visualizing}. As we continue with a very particular field, without direct examples in renal computational pathology, we will analyze an example of activation maximization in brain pathology. In this specific case, the filters of layer 159 that most highly activated the neurons for each class were visualized. When analyzing the filters, features were found that corresponded to the type of morphology found\cite{hollon2020near}.

\subsubsection{Feature visualization}\label{featu_vis}

This research area encompasses a set of techniques and protocols for visualizing features in neural networks. There are a variety of approaches, the main one being optimization to visualize features that can be at the neuron level, at the channel level or at the layer level. Thanks to the optimization, the differences between instances can be interpreted so that two neurons fire together or separately. However, a great challenge in this area, and which extends to all fields of neural networks, is the separation of noise from task-competent information. Especially in feature visualization, we can find representations that are basically noise but that cause neurons to fire in a pattern similar to another image. Being able to identify this and generate representations that make sense is key to being able to interpret an output\cite{olah2017feature}.

\subsubsection{Network dissection}\label{net_diss}

This method basically focuses on understanding the behavior of each layer of a neural network. A dataset is used that has different images where each pixel has a particular label (a mask). For example, if we have an image of a garden, all the pixels that correspond to a flower will have a value of 1 and the rest will have a value of 0. Then the activation map of the layer that we want to study is extracted, taking into account that since it is smaller than the input image, we will have to resize it with bilinear interpolation. When we have the activation map with the same size as the input, we compare it with the mask that we generated first. This way we will know what that layer is focusing on. For example, if we have an image with 3 objects (a plate, a glass and a fork), we will have 3 masks. The plate label mask will have values of 1 in all pixels that overlap with the plate. In this way, when we pass the image through the network and obtain the activation map, if it matches better with the plate mask, we will know what is important for the network\cite{bau2017network}.

\subsection{Applications in renal computational pathology}

Given that the third dimension of interpretability is a relatively modern field and its application tells us more about the model than about the model's task, there are few works applied to Renal computational pathology. One of the works has delved into the features that the model sees when it performs a classification. In addition to having the map of what the model observes, a descriptor has also been incorporated that allows generating a text output on the region of interest (ROI)\cite{zhang2019pathologist}. The images used have been of H\&E histology for urothelial cancer, with a color quite similar to the tumor images of renal samples.

\section{Final considerations}\label{conclu}

In the different sections of this article we briefly saw the different partners of renal computational pathology. Starting with the methods for generating images from tissues, we have briefly explored each of them. We have observed the importance of electron microscopy in the diagnosis of renal pathologies. Likewise, we have investigated a relatively new technique such as hyperspectral imaging to further refine the diagnosis of certain pathologies.

In advancing computational pathology methods in nephrology, we have classified the field of computer vision applications. From that classification, we have briefly explored the field of what is published in the area. We have provided an analysis that includes the medical conditions studied, the algorithms, the number of patients and the data augmentation techniques, among other important parameters. We have observed that most of the applications identify primarily the glomerulus. They continue from there in the analysis of that functional unit to detect pathology. Additionally, we have witnessed the possibility of generating new stainings from H\&E images using generative adversarial networks (GAN).

As an operational part of the pipelines for computational pathology, we have investigated the techniques for aligning consecutive images (multiple stains) of the same surgical piece. This is extremely useful to speed up the identification of structures in different techniques. For example, correlate IHC markers on H\&E images and generate normal and pathological patches.

We have also briefly analyzed the different data protection constraints that different works considered when operating with sensitive patient information. In this particular section we have observed that there is no homogeneous protocol to be followed globally and that many times it is the authors themselves who decide not to submit the research protocol to the review of an institutional ethics committee.

 Finally, we have delved into the dimensions of interpretability and its importance in translating the outputs of our models into clinical medical language. Of these dimensions, we have seen how there is a lack of representativeness of the third dimension (model behaviour) not only in renal computational pathology but also in the general computational pathology of the third dimension. Although applications have been exposed in the general field of computer vision, we have found very few works in computational pathology. This could represent an interesting field of research in the future, since it would propose new hypotheses to illuminate the black box that many models represent.









\section*{Acknowledgment}

The author would like to thank Ramiro Gilardino, MD for reviewing the manuscript.

\bibliography{ref.bib}

\bibliographystyle{IEEEtran}

\vspace{12pt}

\appendix

In this appendix we include a table with a brief cataloging and analysis of the relevant publications in Renal computational pathology. The analysis extracts the objective (use) of the algorithms, the medical conditions studied, the type of algorithm, the type of histological staining (image type), the number of patients included in the study, the magnification levels of the images that fed the models, the data augmentation techniques used, the place of origin of the data (data), the geographical location of the institution that developed the work and the bibliographic reference (Table \ref{Big_table}).

\begin{landscape}

\begin{table}[]
\centering
\caption{Publication analysis of relevant works in Renal computational pathology. GLR, glomeruli; TU, tubule; AR, artery; FB, fibrosis; RO, rotation; Fl, flip; CON, contrast; CR, crop; COL, color; SCA, scaling; BR, brightness; TRA, transpose; NOI, noise; ELA, elastic deformation; DS, digital staining as augmentation; HUE, hue color space transformation}
\resizebox{1.3\textwidth}{!}{%

\begin{tabular}{llllllllllll} 
\hline
\textbf{Use}                      & \textbf{Conditions}                                                                                                                        & \textbf{Algorithm} & \textbf{Image type}     & \textbf{Patients} & \textbf{Magnification (x)} & \textbf{Augmentations}                    & \textbf{Data}                                                                        & \textbf{Location} & \textbf{Year} & \textbf{Reference}                                     &   \\ 
\cline{1-11}
GLR detection                     & Transplant                                                                                                                                 & YOLOv3             & HE                      & 10                &                            & RO, FL, CON                               & Hannover Medical School                                                & Germany, France   & 2020          & \cite{heckenauer2020real}             &   \\
GLR detection                     & N/A                                                                                                                                        & faster R-CNN       & HE, PAS                 & N/A               &  200, 400              & N/A                                       & Taichung Veterans Hospital                                             & Taiwan            & 2018          & \cite{lo2018glomerulus }              &   \\
Condition detection               & Diabetic nephropathy                                                                                                                       & CNN                & IF                      & 83                & N/A                        & N/A                                       & Okayama University                                                     & Japan             & 2020          & \cite{ kitamura2020deep}              &   \\
GLR segmentation                  & Chronic kidney disease                                                                                                                     & Inception v3 CNN   & Masson                  & 171               & N/A                        & NOI                                       & Boston Medical Center                                                   & United States     & 2019          & \cite{ kannan2019segmentation}        &   \\
Condition detection               & Global, segmental sclerosis,                                                                                         & InceptionV3        & PAS, PAM                & 283               & N/A                        & N/A                                       & Kyoto University Hospital                                           & Japan             & 2020          & \cite{ uchino2020classification}      &   \\
GLR segmentation                  & Multiple kidney disease                                                                                                                    & Cascade Mask R-CNN & HE, PAS, PAM, Masson    & 664               & N/A                        & RO, FL, CR, TRA                           & Peking University & China             & 2021          & \cite{ jiang2021deep}                 &   \\
GLR segmentation                  & Transplant                                                                                                                                 & U-Net              & PAS                     & 101               & 200                        & RO, FL, ELA , SCA, CR, NOI, blur, CON, BR & Radboud  Medical Center                                                    & Netherlands       & 2019          & \cite{ hermsen2019deep}               &   \\
GLR, TU, AR segmentation          & Tubulointerstitial nephritis                                                                                                               & U-Net              & PAS                     & 21                & N/A                        & RO, FL, CON                               & Kanazawa  Hospital                                                         & Japan             & 2022          & \cite{ hara2022evaluating}            &   \\
Stain generation                  & Non tumoral renal disease                                                                                                                  & GAN                & HE, PAS, Masson, Jones  & 58                & 200                        & DS                                        & UCLA                                          & United States     & 2021          & \cite{ de2021deep}                    &   \\
Lesion detection                  & Ig A nephropathy                                                                                                                           & LSTM-GCNet, V-Net  & PAS                     & 400               & 400                        & N/A                                       & Jinling Hospital                                                       & China             & 2020          & \cite{zeng2020identification }        &   \\
GLR, TU, AR, FB segmentation      & Transplant                                                                                                                                 & CNN, U-Net         & PAS                     & 15                & 200                        & RO, TRA, ELA                              & Radboud  Medical Center,                                                   & Netherlands       & 2018          & \cite{ de2018automatic}               &   \\
GLR detecion                      & Glomerulosclerosis                                                                                                                         & CNN                & HE                      & 17                & 200                        & FL, RO, TRA                               & Washington University                                     & United States     & 2018          & \cite{ marsh2018deep}                 &   \\
Cancer cell detection             & Renal cell carcinoma                                                                                                                       & CNN                & HE                      & 2093              & N/A                        & FL, CR                                    & The Cancer Genome Atlas                                                              & Global            & 2019          & \cite{ tabibu2019pan}                 &   \\
Cancer cell detection             & Renal cell carcinoma                                                                                                                       & CNN                & HE                      & 42                & 200                        & N/A                                       & The Cancer Genome Atlas                                                              & United States     & 2020          & \cite{fenstermaker2020development }   &   \\
GLR segmentation                  & N/A                                                                                                                                        & U-Net              & HE, PAS, Masson, Silver & 125               & 400                        & N/A                                       & Neptune repository                                                                   & United States     & 2021          & \cite{jayapandian2021development }    &   \\
GLR detection                     & N/A                                                                                                                                        & CNN, HOG           & HE, PAS, Jones          & 26                & 400                        & N/A                                       & Hannover Medical School      & Germany, France   & 2017          & \cite{temerinac2017detection }        &   \\
FB detection                      & FB                                                                                                                                         & glpathnet          & Masson                  & 64                & 400                        & N/A                                       &  Wexner Medical Center                                      & United States     & 2021          & \cite{ zheng2021deep}                 &   \\
GLR detection and classification  & Membranous nephropathy                                                                                                                     & YOLOv3, MIL        & PAM                     & 1281              & 400                        & N/A                                       &  Shanxi Medical University                                         & China             & 2020          & \cite{hao2020classification }         &   \\
Stain generation                  & Transplant                                                                                                                                 & GAN                & PAS, IF                 & 20                & N/A                        & N/A                                       & N/A                                                                                  & N/A               & 2021          & \cite{jen2021silico }                 &   \\
Lesion detection                  & Transplant                                                                                                                                 & CNN                & HE, PAS, Jones          & 2282              & N/A                        & N/A                                       & Netherland hospitals                                                 & Netherlands       & 2022          & \cite{ kers2022deep}                  &   \\
GLR detection and classification  & Lupus nephritis                                                                                                                            & YOLOv4             & PAS                     & 163               & N/A                        & FL, RO, SCA, TRA, HUE, NOI                & Xijing Hospital                                                                      & China             & 2021          & \cite{ zheng2021deep}                 &   \\
GLR detection                     & Transplant                                                                                                                                 & Faster R-CNN       & N/A                     & 19                & N/A                        & RO                                        & Bari University Hospital                                                             & Italy             & 2020          & \cite{altini2020deep }                &   \\
GLR and TU segmentation           & Glomerulosclerosis                                                                                                                         & U-Net              & PAS                     & 83                & 100                        & FL, RO, SCA                               & Citta della Salute                                              & Italy             & 2021          & \cite{salvi2021automated }            &   \\
GLR detection                     & Glomerulosclerosis                                                                                                                         & VGG16              & HE                      & 84                & 200                        & FL, RO                                    & Washington University                                                                & United States     & 2021          & \cite{ marsh2021development}          &   \\
Macrophage, lymphocite detection  & Transplant                                                                                                                                 & U-Net              & IHC                     & 22                & N/A                        & N/A                                       & Hannover Medical School                                                              & Germany           & 2021          & \cite{hermsen2021quantitative }       &   \\
Condition detection               & amyloidosis, diabetic glomerulosclerosis & N/A                & EM                      & 600               & N/A                        & N/A                                       & Northwell Health                                                                     & United States     & 2021          & \cite{ hacking2021deep}               &   \\
Interstitium, TUs segmentation    & Transplant                                                                                                                                 & R-CNN, U-Net       & PAS                     & 93                & 200                        & N/A                                       & N/A                                              & United States     & 2022          & \cite{yi2022deep }                    &   \\
GLR segmentation                  & Normal tissue                                                                                                                              & Mask-RCNN          & IF, PAS                 & N/A               & N/A                        & N/A                                       & N/A                                                                                  & United States     & 2021          & \cite{ patterson2021autofluorescence} &   \\
Cancer cell detection             & Renal Cell Carcinoma                                                                                                                       & CNN                & HE                      & 44                & 400                        & RO, FL, SCA                               & Louisville Hospital                                                        & United States     & 2022          & \cite{abdeltawab2022deep }            &   \\
Deposition detection              & Membranous nephropathy                                                                                                                     & CNN                & Hyperspectral           & 20                & N/A                        & N/A                                       & China Friendship Hospital                                                      & China             & 2021          & \cite{ tu2021deep}                    &   \\
Peritubular capillaries detection & Transplant                                                                                                                                 & CNN                & IHC                     & 380               & 200                        & RO, FL, SCA                               & Asan Medical Center                                                                  & South Korea       & 2019          & \cite{ kim2019fully}                  &   \\
GLR detection                     & N/A                                                                                                                                        & AlexNet            & HE, PAS, Silver, Masson & 28                & 400                        & N/A                                       & Dartmouth Medical Center                                                   & United States     & 2020          & \cite{wilbur2020using}                &   \\
Vascularity detection             & Renal cell carcinoma                                                                                                                       & SVM                & HE, IHC                 & 8                 & 400                        & N/A                                       & Cedars Medical Center                                                          & United States     & 2017          & \cite{ huang2017novel}                &   \\
Cancer cell detection             & Renal cell carcinoma                                                                                                                       & ResNet             & HE                      & 486               & 200                        & FL, RO, COL                               & Dartmouth Medical Center                                                   & United States     & 2021          & \cite{zhu2021development }            &   \\
Glomerular deposits               & N/A                                                                                                                                        & ResNet             & IF                      & 2542              & 400                        & FL, RO                                    & University of Modena,                                              & Italy             & 2020          & \cite{ligabue2020evaluation }         &   \\
Lesion detection                  & IgA, membranous, diabetic nephropathy                                                                              & AlexNet            & IF                      & 1608              & N/A                        & N/A                                       & People’s Liberation Hospital                                            & China             & 2021          & \cite{ pan2021multi}                  &   \\
Podocyte segmentation             &  Antibody–associated glomerulonephritis                                                                          & U-Net              & IF                      & 317               & N/A                        & N/A                                       & Eschweiler Medical Center                                                            & Germany           & 2021          & \cite{zimmermann2021deep}             &   \\
Lesion detection                  & Podocytopathy                                                                                                                              & VGG19              & HE, PAS, PAM            & 835               &                            & FL, RO, BR, SCA                           & Goncalo Moniz Institute                                                              & Brasil            & 2022          & \cite{barros2022podnet}              &   \\
\hline
\end{tabular}
\label{Big_table}

}

\end{table}
 
\end{landscape}

\end{document}